\documentstyle{article}

\newcommand{\be}{\begin{equation}}
\newcommand{\ee}{\end{equation}}
\newcommand{\ben}{\begin{eqnarray}}
\newcommand{\een}{\end{eqnarray}}

\begin{document}

\title{\bf Scalar perturbations and the possible
self-destruction of the {\it phantom menace}}

\author{J.C. Fabris\footnote{Electronic address:
fabris@cce.ufes.br} and
S.V.B. Gon\c{c}alves\footnote{Electronic address:
sergio@cce.ufes.br}\\
Departamento de F\'{\i}sica
Universidade Federal do Esp\'{\i}rito Santo\\
CEP29060-900 Vit\'oria, Esp\'{\i}rito Santo, Brazil}

\date{\today}

\maketitle

\begin{abstract}
Some analysis of the supernovae type Ia observational data seems to indicate that
the Universe today is dominated by a phantom field, for which all energy conditions
are violated. Such phantom field may imply a singularity in a future finite time,
called big rip. Studying the evolution of scalar perturbations for such a field, we
show that if the pressure is negative enough, the Universe can become highly inhomogeneous
and this {\it phantom menace} may be avoided.
\end{abstract}

Using the supernova type Ia as standard candle to evaluate the
distance versus redshift relation for high values of the redshift,
two observational projects arrived at the conclusion that the
Universe today is in a state of accelerated expansion, in
opposition to what was believed until almost one decade ago
\cite{SN}. This accelerated expansion can be achieved if the
Universe today is dominated by a fluid with negative pressure,
that remains a smooth component, not suffering local
agglomeration, as it happens with ordinary matter. This fluid can
be represented by an equation of state $p = w\rho$, with $w < -
1/3$, in order the strong energy condition to be violated, leading
to a negative deceleration parameter $q = - a\ddot a/\dot a^2$,
where $a$ is the scale factor describing the evolution of a
homogeneous and isotropic Universe.
\par
Initially, the SNe type Ia data leads to the estimation that $w
\sim - 0.8$, but the cosmological constant case ($w = - 1$) was
not excluded. Using a sample of 230 SN type Ia, and the hypothesis
of a flat Universe, the authors of reference \cite{tonry} found
$-1.48 < w < -0.72$ at $95\%$ confidence level. With a more
restricted, and selected, sample of 176 SN type Ia, it was found
in reference \cite{riessb} that $w = -1.02_{-0.19}^{+0.13}$. The
precise value of the parameter $w$ is one of the most important
task in observational cosmology today. The evaluation of $w$ from
the observational data, seems to depend not only on the background
model, but also on the sample of data and on the way the analysis
is performed. In any case, there claims that data favors $w < - 1$
\cite{bigrip,hannestad,star03,chandra}, indicating that the
Universe today is dominated by an exotic fluid for which all
energy conditions are violated, denominated for this reason
"phantom fluid". For this fluid, density increases as the Universe
expands, until reaching a singularity (curvature and density
diverge) in a future finite time. The expression "big rip" has
being coined to name this {\it phantom
menace}\cite{bigrip,caldwell}. There are, on the other hand,
claims that this phantom menace may not exist \cite{padmana1}.
\par
In general, when the pressure is negative the fluctuations of the
fluid decreases at large scales, and oscillates with decreasing
amplitude in small scale\cite{jerome,nazira}. Hence, fluids with
negative pressure tend to remain a smooth component of the
Universe. Moreover, when a phantom field dominates the matter
content of the Universe, it tends to damp the evolution of the
matter perturbations \cite{amendola}. The goal of the present
report is to show that, in spite of the previous remarks, when
pressure is negative enough, such that $w \leq - 5/3$,
fluctuations may grow at large scales, and oscillates with
increasing amplitude at small scales. If this happens, when the
phantom fluid dominates, the Universe becomes more and more
inhomogeneous, and in this sense all scenario leading to the big
rip would cease to exist.
\par
In order to verify this, let us consider, for simplicity, a Universe dominated by a fluid
with an equation of state given by $p = w\rho$.  Using the Bardeen's gauge invariant
formalism for the evolution of perturbations, the scalar perturbations for a flat Universe
are essentially represented by the potential $\Phi$, obeying the equation \cite{brand},
\begin{equation}
\Phi'' + 3(1 + w)H\Phi'+ \biggr\{wk^2 + 2H'+ (1 + 3w)H^2 \biggl\}\Phi = 0 \quad ,
\end{equation}
where $k$ is the wavenumber of the perturbation, and $H = a'/a$ is the Hubble parameter measured
using the conformal time $\eta$. Primes mean derivatives with respect to this conformal time.
The scale factor behaves as $a = a_0|\eta|^{2/(1 + 3w)}$. Notice that when $w > - 1/3$, $\eta \rightarrow \infty$ implies $a \rightarrow \infty$, while for $w < - 1/3$, $\eta \rightarrow 0_-$ implies $a \rightarrow \infty$.
\par
Using the solution for the scale factor in terms of the conformal time, the equation
for the potential $\Phi$ becomes,
\begin{equation}
\Phi'' + 6\frac{1 + w}{1 + 3w}\frac{\Phi'}{\eta} + wk^2\Phi = 0 \quad ,
\end{equation}
with the following solutions:
\begin{eqnarray}
\Phi &=& \eta^{-\nu}\biggr\{c_1J_\nu(\sqrt{w}k\eta) + c_2J_{-\nu}(\sqrt{w}k\eta)\biggl\}
\quad , \quad w > 0\quad ; \\
\Phi &=& \eta^{-\nu}\biggr\{c_1I_\nu(\sqrt{w}k\eta) + c_2K_{-\nu}(\sqrt{w}k\eta)\biggl\}
\quad , \quad w < 0\quad .
\end{eqnarray}
In these expressions, $J_\nu$ is the ordinary Bessel function, $I_\nu$ and $K_\nu$ are
modified Bessel functions, the $c$'s are integration constants and $\nu = (5 + 3w)/[2(1 + 3w)]$.
In the large scale limit ($k \rightarrow 0$), for any value of
$w$, the solutions behave as
\begin{equation}
\Phi \sim \tilde c_1 + \tilde c_2 \eta^{-2\nu} \quad ,
\end{equation}
where the $\tilde c$'s are redefined integration constants. There is always a constant mode,
represented by $\tilde c_1$. To analyze the other mode, we must keep in mind that
an expanding Universe implies $\eta \rightarrow \infty$, for $w > - 1/3$, and $\eta \rightarrow
0_-$, for $w < - 1/3$. It is easy to verify that the second mode is always decreasing
if $w > - 5/3$; it grows, however, if $w < - 5/3$. Hence, if the fluid is such
that it violates all energy condition (characterizing a phantom fluid) and if the pressure
is negative enough ($w < - 5/3$), fluctuations grow at large scale, and we can expect that homogeneity
will be destroyed not only locally, as the standard scenario of structure formation requires, but also globally.
\par
At small scales ($k \rightarrow \infty$), the behaviour depends on $w$. For positive $w$,
the potential exhibits decreasing oscillations. For negative $w$, due to the small value
of argument of the modified Bessel functions, we find instabilities. However, these
instabilities occur only for $0 > w > - 1/3$ and $w < - 1$, otherwise a regular behaviour
can be recovered at small scales.
\par
The instabilities at small scales must not be taken so seriously, since it must be due mainly to
the hydrodynamical approximation \cite{nazira}. It is possible to find a more fundamental representation
for such exotic fluids, by considering a self-interacting scalar field, which reproduces,
from the point of view of the background behaviour, the hydrodynamical approach employed until
now. A scale factor which evolves as $a \propto \eta^\frac{2}{1 + 3w}$, with $w < - 1$, can be achieved by
considering a self-interacting minimally coupled scalar field, such that,
\begin{equation}
V(\phi) = V_0\exp\biggr(\pm\sqrt{-3(1 + w)}\phi\biggl)  \quad ,
\quad \phi = \pm 2\frac{\sqrt{-3(1 + w)}}{1 + 3w}\ln\eta \quad .
\end{equation}
\par
For gravity minimally coupled to a (self-interacting) scalar field, the equation for the Bardeen's potential is \cite{brand}
\begin{equation}
\Phi'' + 2\biggr\{H - \frac{\phi''}{\phi'}\biggl\}\Phi' + \biggr\{k^2 + 2\biggr[H'- H\frac{\phi''}{\phi'}\biggl]\biggl\}\Phi = 0 \quad .
\end{equation}
Using the background expressions for $H$ and $\phi$, this equation becomes,
\begin{equation}
\Phi'' + 2 \frac{3(1 + w)}{1 + 3w}\frac{\Phi'}{\eta} + k^2\Phi = 0 \quad ,
\end{equation}
with the solutions
\begin{equation}
\Phi = \eta^{-\nu}\biggr\{c_1J_\nu(k\eta) + c_2J_{-\nu}(k\eta)\biggl\} \quad , \quad
\forall\, w \quad .
\end{equation}
In the large scale limit the solutions behave as in before, in the hydrodynamical representation, and there is still a growing mode when $w \leq - 5/3$. Notice that
the divergence is logarithmic for $w = - 5/3$.
In the small scale limit, however, the potential behaves as
\begin{equation}
\Phi \sim \eta^{-12\nu}\cos(k\eta + \delta) \quad ,
\end{equation}
$\delta$ being a phase. It is easy to verify that for $w > - 1$, the potential oscillates
with decreasing amplitude, while for $w < - 1$, the potential behaves with increasing
amplitude. Hence, in this situation, the phantom field may exhibit instability at
large and small scales. However, it must be stressed that the behaviour at small scale is
quite model-dependent, and another field representation of the phantom field can
modify the conclusions at small scales, like considering the phantom field as a ghost condensation \cite{piazza} or a tachyon \cite{padma2}. But, at large scales, it seems
that there is
always a growing mode, for $w \leq - 5/3$, irrespective of the representation chosen.
\par
Hence, a phantom fluid obeying a barotropic equation of state with $w \leq - 5/3$ is
unstable gravitationally at all scales, and may lead
to a inhomogeneous Universe. This may allow
to avoid the big rip, which is obtained on the condition of homogeneity and isotropy. 

{\bf Acknowledgements}\\
We thank Nelson Pinto-Neto for many discussions. This work has
been partially financed by CNPq (Brazil). \ \\


\begin{thebibliography}{99}

\bibitem{SN}
Riess, A.G. et al., Astron. J. 116, 1009 (1998); Perlmutter, S. et
al., Astrophys. J. 517, 565 (1999); Riess, A.G. et al., Astrophys.
J. 607, 665 (2004).

\bibitem{tonry} J.L. Tonry et al, Astrophys. J. {\bf 594}, 1(2003).

\bibitem{riessb} A.G. Riess, Astrophys. J. {\bf 607}, 665(2004).

\bibitem{bigrip}  Robert R. Caldwell, Marc Kamionkowski, Nevin N. Weinberg, Phys. Rev. Lett. {\bf 91}, 071301(2003).

\bibitem{hannestad}
    S. Hannestad and E. Mortsell, {\it JCAP\/} {\bf 0409}, 001 (2004).

\bibitem{star03}
    U. Alam, V. Sahni, T.D. Saini and A.A. Starobinsky,
        {\it Mon. Not. R. Astron. Soc.\/} {\bf 354}, 275 (2004).

\bibitem{chandra}
    S.W. Allen et al.,
    {\it Mon. Not. R. Astron. Soc.\/} {\bf 353}, 457 (2004).

\bibitem{caldwell} R.R. Caldwell, Phys. Lett. {\bf B545}, 23(2002).

\bibitem{padmana1} H.K. Jassal, J.S. Bagla and T. Padmanabhan, {\it The vanishing
phantom menace}, astro-ph/0601389.

\bibitem{jerome} J.C. Fabris and J. Martin, Phys. Rev. {\bf D55}, 5205(1999).

\bibitem{nazira} J.C. Fabris, S.V.B. Gon\c{c}alves and N.A. Tomimura, Class. Quant. Grav.
{\bf 17}, 2983(2000).

\bibitem{amendola} L. Amendola, S. Tsujikawa and M. Sami,  Phys. Lett. {\bf B632}, 155(2006).

\bibitem{brand} V.F. Mukhanov, H.A. Feldman and R.H. Brandenberger, Phys. Rep. {\bf 215}, 203(1992).

\bibitem{piazza} F. Piazza and S. Tsujikawa, JCAP {\bf 0407}, 004(2004).

\bibitem{padma2} J.S. Bagla, H.K. Jassal and T. Padmanabhan, Phys. Rev. {\bf 67}, 063504(2003).
\end{thebibliography}
\end{document}